\documentclass[
   ,final            
  ]
  {aipproc}

        \def\dis{\displaystyle}
  \def\be{\begin{equation}}
  \def\m{\multicolumn}    \def\ra{\rightarrow}
  \def\beq{\begin{eqnarray}}  \def\eeq{\end{eqnarray}}
  \def\be{\begin{equation}}   \def\ee{\end{equation}}

\layoutstyle{6x9}
\usepackage{hyperref}
\begin{document}

\title{Masses and decay modes of charmonia using \\ a confinement model}


\author{J N Pandya}{
  address={Department of Physics, Veer Narmad South Gujarat University, Surat 395 007, INDIA.},
  }

\author{Ajay Kumar Rai}{
  address={Applied Physics Department, Faculty of Technology \& Engineering,\\ The M S University of Baroda, Vadodara 390 001, INDIA.},
}

\author{P C Vinodkumar}{
  address={Department of Physics, Sardar Patel University, Vallabh Vidyanagar 388 120, INDIA.},
}

\begin{abstract}
 The masses of charmonium $s$ and $p$-states, pseudoscalar and vector decay constants, leptonic, hadronic  as well  as radiative decay widths for charmonia have been computed in the framework of extended harmonic confinement model without any additional parameters. The outcome in comparison with other contemporary theoretical and experimental results is presented.
\end{abstract}

\maketitle


  The experimental investigations in charmed quark sector has opened up
  a vast data bank for  the hadronic decays into various channels
  \cite{pdg2006,cleo2006,wys2005}. A hadronic transition or decay between quarkonium levels is understood as a two-step process in which gluons first are emitted from the heavy
  quarks annihilating each other and then recombine into light hadrons. Perturbative QCD is not directly applicable because the energy available to the light hadrons is small and the emitted gluons are
  soft. Nevertheless, the final quarkonium state is small compared to the system of light hadrons and moves nonrelativistically in the rest frame of the decaying quarkonium state.

  Decay rates of heavy quarkonium states in the non-relativistic
  limit with some QCD radiative correction have been studied
  through various phenomenological models subjected to
  substantial relativistic corrections \cite{ls2006}.
  In the heavy flavour sector, the nonrelativistic
   expressions in accordance with the factorizable
   assumption for the long-distance part and the
   short-distance part of the decay amplitudes,
   are proportional to the square of the radial
   wave function at the origin.  We employed there a relativistic
harmonic confinement model with scalar plus vector  potential
\cite{pcvjnp1999}.
  \section{Hadronic masses from ERHM}
   The mass of a hadron having $p$ number of quarks in ERHM can be obtained as \cite{pcvjnp1999},
  \be
  M_N (q_1q_2.....)\ =\ \dis\sum_{i=1}^{p}\epsilon_N(q_i,p)_{conf}\ +\
  \dis\sum_{i<j=1}^{p}\epsilon_N (q_iq_j)_{coul} + \dis\sum_{i<j=1}^p \epsilon_N^J(q_i,q_j)_{SD}    \label{eq:hadronmass}\ee
  where the first sum is the total confined energies of the constituting quarks of the hadron, the second sum corresponds to the residual colour coulomb interaction energy between the confined quarks and the third sum is due to spin dependent terms.

  The intrinsic energies of the quarks in charmonium system is given by
  \be
  \epsilon(N)_{conf}\ = \left [(2N+3)\Omega_N (q)\ +\ M_q^2\ -\ 3
  M_q \Omega_0(q)/{\dis\sum_{i=1}^2 M_q(i)}\right ]^{1/2} \label{eq:epsilon} \ee

  The coulombic part of the energy is computed using the residual coulomb potential given by \cite{gw1986},  \be V_{coul}(q_iq_j)\ =\ \dis\frac{k \ \alpha_s(\mu)}{\omega_n r} \ee
  here $\omega_n$ represents the colour dielectric ``coefficient'' \cite{gw1986}.   $\omega_n$ is found to be to be state dependent \cite{pcvjnp1999}, so as to get consistent coulombic contribution to the excited states of the hadrons. It is a measure of the confinement strength through the nonperturbative contributions to the confinement scale at the respective threshold energy of the given flavour production.

  We construct the wave functions for charmonia by  retaining the nature of single particle wave function but with a two particle size parameter $\Omega_N({q_iq_j})$ instead of $\Omega_N(q)$ \cite{jnppcv2001}.  Now coulomb energy is computed perturbatively using the confinement basis with two particle size parameter defined above for different  states as,
  $\epsilon_N(q_iq_j)_{coul}\ =\ \langle N | V_{coul} | N \rangle$.
  The fitted parameters to obtain experimental ground state mass are $m_c$ = 1428 MeV, $k = 0.19252$ and
  the confinement parameter $A = 2166$ MeV$^{3/2}$.

  The confined gluon propagator derived in the current confinement scheme (CCM) for gluons has been employed in the derivation of one gluon exchange potential using RHM  wave functions for quarks under the Breit formalism \cite{vvk1992}. This COGEP has central, tensor and  spin-orbit terms in addition to many momentum dependent terms.

  From the centre of weight masses, the pseudoscalar and vector mesonic masses are computed by incorporating the residual two body chromomagnetic interaction through the spin-dependent term of the COGEP perturbatively as,
  \be \epsilon_N^J(q_iq_j)_{S.D.}\ =\ \langle NJ | V_{SD}|NJ \rangle \ee
  where $|NJ\rangle$ is the given hadronic state. For mesons $|NJ\rangle$ becomes the $|q_i\bar q_j\rangle$ states.  We consider the two body spin-hyperfine interaction of the residual (effective) confined one gluon exchange potential (COGEP) as \cite{vvk1992},
  \be
  V_{\sigma_i\cdot \sigma_j}\ =\ {\dis\frac{\alpha_s(\mu) N_i^2N_j^2}{4}}
  \dis\frac{\lambda_i\cdot \lambda_j}{[E_i + m_i] [E_j + m_j]}
  [4\pi \delta^3(r_{ij}) - C_{_{CCM}}^4 r^2
  D_1(r_{_{ij}})]\left(-\dis\frac{2}{3}{\sigma_i\cdot \sigma_j}\right)
  \label{eq:ssintpot}\ee
  The spin-orbit interaction of the residual (effective) confined one gluon exchange potential (COGEP) is given as \cite{vvk1992}
  \be V_{q_iq_j}^{LS} = \dis\frac{\alpha_s}{4}\cdot\dis\frac{N_i^2\
  N_j^2}{(E_i+M_i)(E_j+M_j)}\dis\frac{\lambda_i\cdot\lambda_j}{2r_{ij}}
  \left [4 \vec L \cdot \vec S \left (D_0'(r_{ij}) +
  2D_1'(r_{ij})\right ) \right ] \label{eq:sointpot1}\ee
  In the case of mesonic states, $\vec L$ and $\vec S$ correspond to
  the orbital and spin operators of the relative co-ordinates. Where $D_0'(r_{ij})$
  and $D_1'(r_{ij})$ appeared in Eqns (\ref{eq:ssintpot}) and (\ref{eq:sointpot1}) are derivatives of the confined gluon propagators of CCM \cite{vvk1992}.

  \section{Decay properties of mesons}
  Without any additional free parameters, using the Van-Royen-Weisskopf formula for computing leptonic decay widths \cite{qr1979} without radiative correction term, we have computed leptonic decay widths and are tabulated in Table \ref{tab:leptdcaywdth} alongwith other theoretical as well as experimental values.

  The Van Royen - Weisskopf formula used for the meson decay constants is obtained in the two-component spinor limit \cite{vrw1967}. $f_P$ and $f_V$ are related to the ground state radial wave function $R_{1S}(0)$ at the origin, by the Van Royen - Weisskopf formula modified for the colour as,
  $ f_{P/V}^2\ = \ 3 |R_{1S}(0)|^2 / \pi M_{P/V}
  \label{eq:pdconst}$  where $M_{P/V}$ is the ground state mass of the pseudoscalar/vector meson. In the present computations, $f_P$ = 410.34 MeV which is closer to the experimental value of 410 $\pm$ 15 MeV, while the other results are 496.42 \cite{qr1979}, 357 \cite{cetal2004}, 349 \cite{zw2005}, 498.90 \cite{bt1981}, 549.61 \cite{mar1980} and 663.06 \cite{eit1978} (all values in MeV).

  We employ radial wave functions to compute the hadronic as well as radiative decay widths of
  the vector and pseudoscalar mesons of $c\bar{c}$ system based on the
treatment of perturbative QCD as \cite{fec1979}
  \beq \Gamma(n ^3S_1 \ra hadron) & = & \dis{\frac{40}{81\pi}} (\pi^2 - 9)
  \alpha_s^3 {\dis\frac{|R_{nS}(0)|^2}{M_n^2}} \\
   \Gamma(n ^1S_0 \rightarrow hadron) & = & \dis{\frac{8}{3}}
   \alpha_s^2 {\dis\frac{|R_{nS}(0)|^2}{M_n^2}}\ ;\ \ \Gamma(n ^1S_0 \rightarrow
   \gamma \gamma)  =  \dis 12 \alpha^2 e_q^4
  {\dis\frac{|R_{nS}(0)|^2}{M_n^2}} \eeq
  Here, symbols have their usual meanings. The computed decay widths in comparison with other theoretical
  predictions are tabulated in Table \ref{tab:plmsgmagama}.

\begin{table}
\caption{Masses of $c\bar c$ in comparison with PDG and other theoretical models}\label{tab:masses}
\begin{tabular}{ccccccccccc}
  \hline
  State  & $J^{PC}$ & Present  & \cite{pdg2006} &\cite{bgs2005}  & \cite{vij2007} & \cite{efg2003} & \cite{bd2003} & \cite{rr2005}pert & \cite{rr2005}nonpert\\
  \hline
  $\eta_c(1^1S_0)$ & $0^{-+}$ & 2985  & 2980 & 2981 & 2990 & 2979 & 3093 & 2980 & 2982\\
  $\eta_c(2^1S_0)$ & $0^{-+}$ & 3626  & 3638 & 3625 & 3627 & 3588 & 3096 & 3597 & 3619\\
  $\eta_c(3^1S_0)$ & $0^{-+}$ & 4047  & --   & 4032 & --   & 3991 & --   & 4014 & 4053\\
  $\eta_c(4^1S_0)$ & $0^{-+}$ & 4424  & --   & 4364 &  --  & --   & --   & --    & -- \\
  \hline
  $J/\psi(1^3S_1)$ & $1^{--}$ & 3096 & 3097 & 3089 & 3097 & 3096 & 3096 & 3097 & 3097\\
  $\psi(2^3S_1)$ & $1^{--}$   & 3690 & 3686 & 3666 & 3685 & 3686 & 3476 & 3686 & 3686\\
  $\psi(3^3S_1)$ & $1^{--}$   & 4082 & 4040 & 4060 & 4050 & 4088 & 3851 & 4095 & 4102\\
  $\psi(4^3S_1)$ & $1^{--}$   & 4408 & 4415 & 4386 & 3443 & --   & 4223 & 4433 & 4447\\
  \hline
  $\chi_{c0}(1^3P_0)$ & $0^{++}$ & 3431 & 3415 & 3425 & 3496 & 3424 & 3468 & 3416 & 3415\\
  $\chi_{c1}(1^3P_1)$ & $1^{++}$ & 3464 & 3511 & 3505 & 3525 & 3510 & 3468 & 3508 & 3511\\
  $\chi_{c2}(1^3P_2)$ & $2^{++}$ & 3530 & 3556 & 3556 & 3507 & 3556 & 3467 & 3558 & 3556\\
  $h_{c1}(1^1P_1)$    & $1^{+-}$ & 3497 & 3526 & 3524 & --   & 3526 & 3467 & 3527 & 3524\\
  \hline
  $\chi_{c0}(2^3P_0)$ & $0^{++}$ & 3891 & 3800 & 3851 & -- & 3854  & 3814 & 3844 & 3864\\
  $\chi_{c1}(2^3P_1)$ & $1^{++}$ & 3899 & 3880 & 3923 & -- & 3929  & 3815 & 3940 & 3950\\
  $\chi_{c2}(2^3P_2)$ & $2^{++}$ & 3916 & 3940 & 3970 & -- & 3972  & 3815 & 3994 & 3992\\
  $h_{c2}(2^1P_1)$    & $1^{+-}$ & 3907 & 3904* & 3941 & -- & 3945 & 3815 & 3961 & 3963\\
  \hline
  \end{tabular}
  \end{table}
 \begin{table}
 \caption{Leptonic decay widths (in keV) of $c\bar c(n^3S_1)$}\label{tab:leptdcaywdth}
 \begin{tabular}{cccccccc}
  \hline
  State & Present & \cite{pdg2006} & \cite{rr2007}pert & \cite{rr2007}nonpert & \cite{qr1979} & \cite{eit1978} & \cite{ir2006} \\
  \hline
  $J/\psi(1^3S_1)$ & 5.469 & 5.55 $\pm$ 0.14 & 4.28 & 1.89 & 4.80 & 7.82 & 6.72 $\pm$ 0.49 \\
  $\psi(2^3S_1)$   & 2.140 & 2.48 $\pm$ 0.06 & 2.25 & 1.04 & 1.73 & 3.83 & 2.66 $\pm$ 0.19 \\
  $\psi(3^3S_1)$   & 0.796 & 0.86 $\pm$ 0.07 & 1.66 & 0.77 & 0.98 & 2.79 & 1.45 $\pm$ 0.07 \\
  $\psi(4^3S_1)$   & 0.288 & 0.58 $\pm$ 0.07 & 1.33 & 0.65 & 0.51 & 2.19 & 0.52 $\pm$ 0.02 \\
  \hline
  \end{tabular}
  \end{table}

\begin{table}
\caption{$\eta_c$ $\rightarrow$ $\gamma \ \gamma $ (in keV) with
correction terms and light hadron decay (in MeV) of $c \bar c$
meson}\label{tab:plmsgmagama}
\begin{tabular}{lccccc}
\hline Models&  \m{1}{c}{$\Gamma$}&
\m{1}{c}{$\Gamma_{NRQCD}$}&\m{1}{c}{$\Gamma_{EXP.}$}&
\m{1}{c}{$\Gamma(a)$}&
\m{1}{c}{$\Gamma(b)$} \\
\hline
ERHM\cite{pcvjnp1999, jnppcv2001} &7.67&&7.06&10.90&0.79 \\
BT\cite{bt1981}&11.19&7.5 \cite{brat2003}&$\pm$&19.55&1.44\\
PL(Martin)\cite{mar1980}&13.81&&0.8&38.20&3.36\\
Log\cite{qr1979}&11.26&9.02 \cite{bod1995}&$\pm$&20.84&1.58\\
Cornell\cite{eit1978}&20.09&&2.3&34.86&2.21\\
$CPP_{\nu=1.4}-VS$\cite{akr2005,akr2002}&8.36&&&13.72&0.84\\
\hline
N Fabiano \cite{fb2002}&&&&14.38$\pm$&\\
&&&& 1.07$\pm 0.55$&\\
\hline
a$ (\eta_c \rightarrow LH)$&b($J/\psi \rightarrow LH$)&&&&\\
\end{tabular}
\end{table}

 \bibliographystyle{aipproc}   

\end{document}